\documentclass[final,nolockup]{fctr}
\usepackage{fctr-fonts}
\newsavebox{\fitb}
\newcommand{\fit}[1]{\sbox{\fitb}{#1}\ifdim\wd\fitb>\linewidth\resizebox{\linewidth}{!}{\usebox{\fitb}}\else\usebox{\fitb}\fi}
\title{VISAI OpenSpeech: An Encoder-Decoder ASR Model\\with Adjustable Inference Cost}

\author{%
  Chompakorn Chaksangchaichot\\[3pt]
  \affil{VISAI AI, Thailand}\\[1pt]
  \email{chompakornc\_pro@vistec.ac.th}\\[3pt]}

\resources{%
  \resourcerow{Weights}{\url{https://huggingface.co/collections/VISAI-AI/visai-open-speech}}
  \resourcerow{Server}{\url{https://github.com/visai-ai/visai-openspeech-inference-server}}}

\begin{document}
\maketitle

\begin{abstract}
We present VISAI OpenSpeech, a 337M-parameter speech recognition model for Thai, English and Chinese, and VISAI OpenSpeech Streaming, its streaming variant. The model combines a FastConformer encoder with a connectionist temporal
classification (CTC) head and an attention decoder, so one checkpoint supports three decoding setups that trade speed
for accuracy at serving time. We train it on 71.5k hours of speech from public corpora and in-house data. On a six-group
Thai benchmark, beam search reaches a weighted character error rate (CER) of 13.48\%, lower than every open model and
all but one of the commercial services we evaluated. On one V100 GPU under load, our server runs at 637 times real time
with greedy CTC (15.23\% CER) and 241 times with beam search. The streaming variant reaches 15.21\% CER with greedy
CTC at 1,040\,ms look-ahead. We release the weights and an ONNX inference server that supports word boosting and WebSocket streaming.
\end{abstract}

\section{Introduction}
End-to-end models have replaced hybrid pipelines in most automatic speech recognition (ASR) systems. Connectionist
temporal classification (CTC) \cite{graves2006ctc} and attention-based encoder-decoder (AED) models \cite{chan2015las}
map audio to text with a single network, and joint CTC/attention training \cite{watanabe2017hybrid} combines the
monotonic alignment of the first with the implicit language model of the second. Conformer \cite{gulati2020conformer} and its faster
variant FastConformer \cite{rekesh2023fastconformer} are widely used as speech encoders. Recent multilingual AED systems
trained on hundreds of thousands of hours reach low error rates \cite{radford2023whisper,sekoyan2025canary,qwen3asr},
but with large parameter counts and slow autoregressive decoding.

\begin{figure}[t]
\centering
\includegraphics[width=\linewidth]{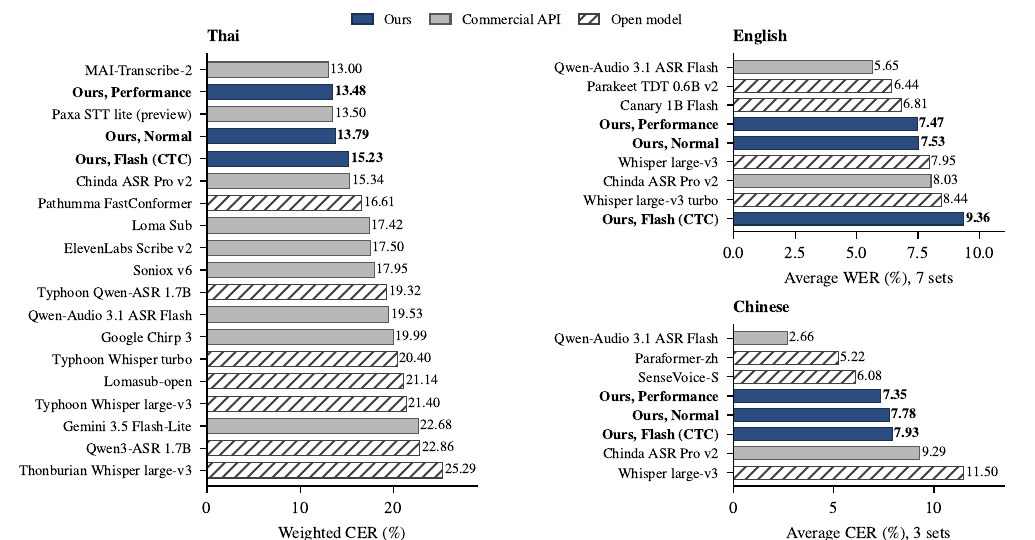}
\caption{Thai weighted CER, English average word error rate (WER) and Chinese average CER (\%) of our setups and other systems (lower is better). All Thai systems, plus Chinda ASR Pro v2 and Qwen-Audio 3.1 ASR Flash on English and Chinese, ran in one pipeline. Other English and Chinese baselines are published results (Tables~\ref{tab:en} and~\ref{tab:zh}). Averages cover the seven English sets and, for Chinese, AISHELL-1 and WenetSpeech test\_net and test\_meeting, as published baselines omit FLEURS.}
\label{fig:thai}
\end{figure}

ASR for Thai is harder than for English in three respects. Thai script does not separate words with spaces, so
recognition errors are measured at the character level. Thai speakers often switch to English within a sentence.
Numbers, dates and units have a spoken form that differs from their written form. Most open Thai ASR models are Whisper
\cite{radford2023whisper} models fine-tuned on Thai data \cite{typhoonwhisper,thonburian}. These models decode
autoregressively, which makes inference slow, and they transcribe only complete utterances. On noisy or long recordings
they can also fall into repetition loops. Commercial ASR services reach lower error rates, but their models and training data
are not available.

The motivation of this work is an open recognizer for Thai, English and Chinese that comes close to commercial
services on Thai and transcribes both recorded files and live audio. We use a 337M-parameter model in which a
247M-parameter FastConformer encoder feeds a CTC head and a 6-layer attention decoder. Both heads read the same encoder
output, so a server selects \emph{Flash} (greedy CTC), \emph{Normal} (greedy decoding) or \emph{Performance} (beam search) per request
without loading a second model. A streaming variant uses a chunked encoder \cite{noroozi2024stateful} and a decoder
that waits for more audio.

On a Thai benchmark of read, broadcast, dialect and meeting speech, VISAI OpenSpeech reaches a weighted CER of 13.48\%
with beam search and 15.23\% with greedy CTC, against 16.61\% for the best open model and 13.00\% for the best
commercial service we evaluated (Figure~\ref{fig:thai}). VISAI OpenSpeech Streaming reaches 15.21\% at 1,040\,ms
look-ahead with CTC and 14.74\% at 320\,ms with its streaming decoder.

This report makes the following contributions:
\begin{enumerate}
\item We present VISAI OpenSpeech, a 337M-parameter model with a FastConformer encoder, a CTC head and an attention
decoder that serves three decoding setups from one checkpoint, and VISAI OpenSpeech Streaming, a chunked variant
trained from it. With beam search the offline model reaches 13.48\% weighted CER on Thai, lower than every open
model we evaluated, and the streaming variant matches offline CTC at 1,040\,ms look-ahead. On one V100 GPU, greedy CTC has 2.6 times the saturated throughput of beam search.
\item We adapt Final Token Preference Optimization (FTPO) \cite{paech2026ftpo}, proposed for repetition loops in language
models, to the attention decoder of an ASR model. A decoder-only fine-tune of 1,000 steps on 1,803 mined loop points
lowers the greedy Thai weighted CER from 15.66\% to 13.79\%, mostly on far-field meeting speech (55.07\% to 44.16\%).
\item We release the weights\footnote{\url{https://huggingface.co/collections/VISAI-AI/visai-open-speech}} and an ONNX inference
server\footnote{\url{https://github.com/visai-ai/visai-openspeech-inference-server}} that runs on CPU or GPU, batches recorded audio across
requests, streams live audio over WebSocket, and boosts per-request phrase lists with a byte-level automaton that
needs no word segmentation for Thai or Chinese. We compare the model on Thai with 13 commercial services and 11 open
models, offline and streaming, in one evaluation pipeline.
\end{enumerate}

Section~2 describes the model, its prompt and its decoding setups. Section~3 covers the training data and the two
training stages. Section~4 describes the inference server. Section~5 reports accuracy and
inference cost for both models and the effect of FTPO on repetition, and Section~6 covers related work.

\section{Model}
\subsection{Encoder}
\begin{figure}[t]
\centering
\begin{minipage}[c]{0.38\linewidth}\centering\small
\fit{\begin{tabular}{lrrrr}
\toprule
Component & Layers & Width & Heads & Params \\
\midrule
Encoder & 18 & 768 & 12 & 247M \\
CTC head & 1 & 768 & & 16M \\
Decoder & 6 & 768 & 12 & 74M \\
\midrule
Total & & & & 337M \\
\bottomrule
\end{tabular}}
\captionof{table}{Model configuration. The CTC head is a single linear layer.}\label{tab:config}

\vspace{10pt}
\footnotesize
\fit{\begin{tabular}{lrrr}
\toprule
Source & Thai & English & Chinese \\
\midrule
GigaSpeech 2 \cite{yang2024gigaspeech2} & 10,004 & & \\
YODAS v2 and v3 \cite{li2024yodas} & 1,967 & 20,627 & 297 \\
GigaSpeech XL and L \cite{chen2021gigaspeech} & & 5,429 & \\
People's Speech \cite{galvez2021peoples} & & 2,265 & \\
SPGISpeech \cite{oneill2021spgispeech} & & 1,580 & \\
VoxPopuli \cite{wang2021voxpopuli} & & 479 & \\
LibriHeavy \cite{kang2024libriheavy} & & 253 & \\
AMI, individual headset \cite{carletta2005ami} & & 73 & \\
WenetSpeech \cite{zhang2022wenetspeech} & & & 9,785 \\
MAGICDATA read speech & & & 588 \\
AliMeeting, near and far \cite{yu2022m2met} & & & 182 \\
AISHELL-1 \cite{bu2017aishell} & & & 151 \\
Other public corpora & 116 & 22 & 459 \\
In-house data & 16,405 & 708 & 121 \\
\midrule
Total & 28,492 & 31,436 & 11,582 \\
\bottomrule
\end{tabular}}
\captionof{table}{Training data after deduplication and test-set decontamination, in hours. Column sums can differ from the
totals because of rounding. For Thai, the other public corpora are LOTUSDIS, the Typhoon Isan corpus and FLEURS. For English, they are FLEURS and CS-Dialogue. For Chinese, they are FLEURS, CS-Dialogue, MagicData-RAMC, AISHELL-4, AISHELL-5, Chinese-LiPS and SeniorTalk.}\label{tab:data}
\end{minipage}\hfill
\begin{minipage}[c]{0.58\linewidth}\centering
\includegraphics[width=\linewidth]{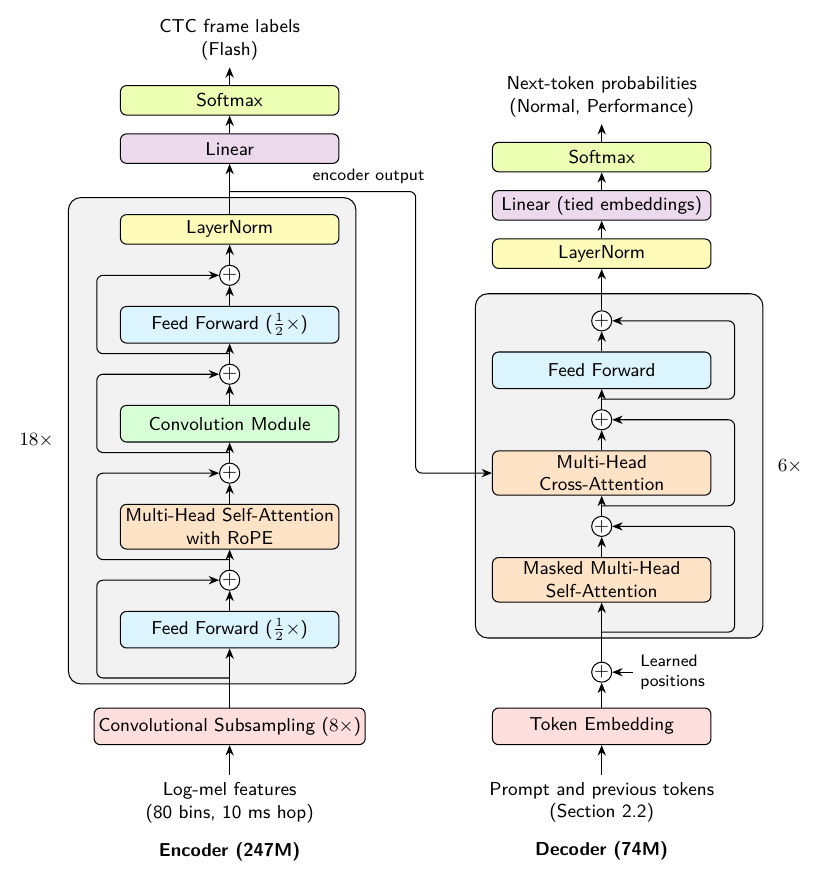}
\caption{VISAI OpenSpeech. A FastConformer encoder (left) feeds a CTC head and an attention decoder (right). The CTC branch gives the \emph{Flash} setup, and the decoder gives \emph{Normal} (greedy) and \emph{Performance} (beam search).}\label{fig:arch}
\end{minipage}
\end{figure}

The encoder follows FastConformer \cite{rekesh2023fastconformer}. Three stride-2 depthwise-separable convolutions
subsample 80-dimensional log-mel features (25\,ms window, 10\,ms hop) by a factor of 8, so one encoder frame covers
80\,ms. The encoder stacks 18 blocks with a model dimension of 768 and 12 attention heads (Figure~\ref{fig:arch},
Table~\ref{tab:config}). Each block follows the Macaron order of Conformer \cite{gulati2020conformer}: a half-step
feed-forward module, multi-head self-attention, a convolution module with kernel size 9, a second half-step
feed-forward module and a final layer normalization. Unlike the standard FastConformer block, the feed-forward modules
use SwiGLU \cite{shazeer2020glu} with an inner dimension of 2,048, self-attention uses rotary position embeddings
\cite{su2021roformer} instead of relative positional encoding, and queries and keys are normalized per head before the
dot product. The encoder has 247M parameters.

\subsection{Heads and prompt}
\begin{figure}[b]
\centering
\includegraphics[width=0.84\linewidth]{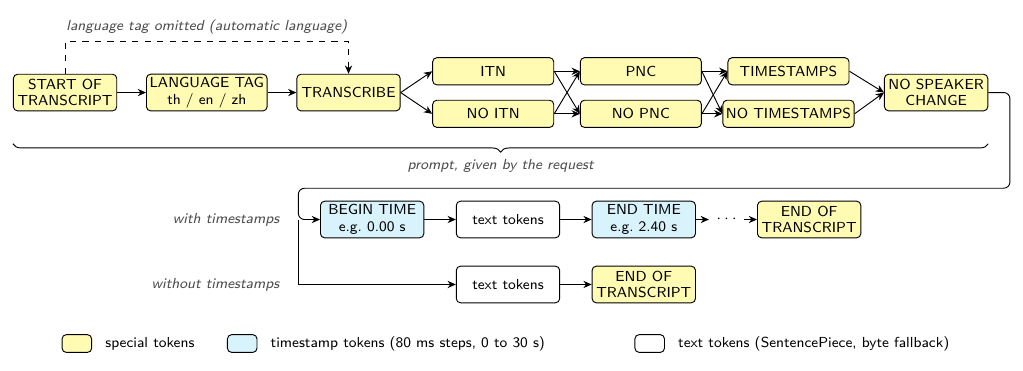}
\caption{Decoder prompt format. The request fixes every prompt token. The language tag is optional. Without it, the decoder transcribes in the language it hears. Timestamp tokens are interleaved with the text only when requested.}
\label{fig:prompt}
\end{figure}

A linear CTC head predicts one of 21,254 labels for every frame: the blank, 256 byte tokens and 20,997 text pieces.
The attention decoder is a 6-layer pre-norm Transformer \cite{vaswani2017attention} with a model dimension of 768,
12 heads, a feed-forward dimension of 3,072 and learned absolute position embeddings. Its input and output embeddings
are tied, giving 74M parameters. Following
Whisper \cite{radford2023whisper}, the decoder is conditioned on a prompt of special tokens (Figure~\ref{fig:prompt}).
The prompt contains a start token, an optional language tag, a task token, and four flags. The first two select
written or spoken forms (ITN, inverse text normalization of numbers, dates and units) and punctuation with case (PNC).
The third requests segment timestamps on an 80\,ms grid. The fourth disables speaker-turn markers, and the server always sends it. Without the language tag, the decoder transcribes in the language it hears. Both heads share one
SentencePiece \cite{kudo2018sentencepiece} unigram vocabulary of 21,885 entries with byte fallback. It joins separately
trained Thai (6,008), Chinese (4,800) and Latin-script (10,149) pieces with 40 common symbols and 888 special, byte and
timestamp tokens.

\subsection{Training objective}
Both heads are trained jointly. The CTC branch sees two views of each batch that differ only in their SpecAugment masks,
and a consistency term \cite{yao2024crctc} penalizes the symmetric KL divergence between the frame posteriors of the two
views. The training loss is
\begin{equation}
\mathcal{L} = 0.3\,\bigl(\mathcal{L}_{\mathrm{CTC}} + \alpha\,\mathcal{L}_{\mathrm{CR}}\bigr) + 0.7\,\mathcal{L}_{\mathrm{CE}},
\end{equation}
where $\mathcal{L}_{\mathrm{CTC}}$ and $\mathcal{L}_{\mathrm{CE}}$ are averaged over the two views. The consistency weight
$\alpha$ rises linearly from 0 to 0.2 over the first 2,000 steps, and $\mathcal{L}_{\mathrm{CE}}$ uses label smoothing
of 0.1.

\subsection{Decoding setups}
The server offers three decoding setups. \emph{Flash} decodes the CTC head greedily in a single encoder
pass with no autoregressive steps, so it cannot fall into repetition loops. \emph{Normal} runs the attention decoder greedily, with at most 448 tokens
per 30\,s window. \emph{Performance} runs a beam search of width 4 with length normalization. All three read the same encoder
output, so one loaded model serves all three. Long recordings are first cut at pauses into windows of at most 30\,s.

\subsection{Streaming variant}\label{sec:streamvar}
VISAI OpenSpeech Streaming keeps the offline architecture but restricts encoder self-attention to chunks. As in cache-aware
FastConformer \cite{noroozi2024stateful}, each frame attends to its own chunk and to 70 cached frames (5.6\,s) of left
context. Chunks of 1, 5 and 14 frames give look-aheads of 0, 320 and 1,040\,ms, and one model serves all three. At
inference the encoder keeps a key-value and convolution cache for each stream and processes one chunk per call, and the
CTC head labels the new frames.

The streaming decoder reads only the encoder frames received so far. After each chunk it emits text tokens until it
predicts a dedicated wait token, and then waits for the next chunk. In a live session the server ends a decoder segment
after 640\,ms without speech in the CTC output, or after 30\,s of audio, and starts a new segment with a fresh prompt.

\section{Data and training}
\subsection{Data}
Table~\ref{tab:data} lists the training data. It contains 71.5k hours of transcribed speech: 28.5k hours of Thai, 31.4k
of English and 11.6k of Chinese. Public corpora provide 54.3k hours (76\%) and in-house data the remaining 17.2k hours.
The largest public sources are GigaSpeech~2 \cite{yang2024gigaspeech2} for Thai, YODAS \cite{li2024yodas} and
GigaSpeech \cite{chen2021gigaspeech} for English, and WenetSpeech \cite{zhang2022wenetspeech} for Chinese. Deduplicating
exact and near-duplicate recordings within and across sources removed about 1.1k hours. We also removed every
training recording that overlaps a test set used in this report. This removed 65 hours of English, mostly LibriSpeech
chapters contained in GigaSpeech audiobooks and SPGISpeech calls that appear in the Earnings-22 test set, as well as
1.4 hours of Chinese and 0.2 hours of Thai.

\subsection{Training}
\label{sec:training}
We pre-train from random initialization for 361,735 steps on 4 NVIDIA A100 40\,GB GPUs in 3.8 days (363 GPU-hours). Each step processes 387\,s of padded audio per GPU in each of the two views. We train with AdamW
\cite{loshchilov2019adamw} ($\beta_1=0.9$, $\beta_2=0.99$, weight decay 0.068) and a warmup-stable-decay schedule with 7,859 warmup steps to a peak learning rate of $5.0\times10^{-4}$, then a square-root decay to zero over the last 20\% of
steps. The maximum window length grows from 15\,s to 30\,s early in training. We use bf16 mixed precision, dropout of
0.1 and adaptive gradient clipping at twice the median recent gradient norm. SpecAugment \cite{park2019specaugment}
applies two frequency masks of up to 27 bins and ten time masks of up to 5\% of the frames. We also add noise from MUSAN
\cite{snyder2015musan}, reverberation, far-field, telephone-band and codec simulation. Thai, English and Chinese are
sampled with weights 0.40, 0.44 and 0.16, and about 5\% of the windows contain no speech. Pre-training sees each language
about 1.85 times. We keep an exponential moving average of the weights with decay 0.9999 and use the averaged model.

Greedy decoding of the pre-trained model sometimes repeats a phrase until it reaches the token limit. We reduce this
with Final Token Preference Optimization (FTPO) \cite{paech2026ftpo}, a short fine-tune of the decoder on mined loop
points. To mine them, we greedily decode a sample of training windows and find 1,803 points where the decoder starts
to repeat itself. Each point gives a preference pair rejecting the first repeated token and choosing the reference token. We then fine-tune the decoder for 1,000 steps with the ASR loss plus the FTPO loss (weight 0.1),
which combines a margin loss that favors the chosen token over the rejected one with a squared-error term that keeps
the remaining logits close to those of the pre-trained model. The encoder and the CTC head stay frozen. The released
offline model is this checkpoint.

The streaming model is trained in two stages from the offline model. The first stage fine-tunes all parameters with
chunked attention for 195,763 steps, one pass over the data, with a peak learning rate of $1.24\times10^{-4}$. The
look-ahead is drawn for each micro-batch: 320\,ms with probability 0.30, 480\,ms 0.20, 160\,ms 0.10, 1,040\,ms 0.10,
0\,ms 0.10, 80\,ms 0.05, and full context 0.15. The second stage freezes the encoder and the CTC head and trains only the
decoder at 320\,ms look-ahead, for another 195,763 steps with a peak learning rate of $6.2\times10^{-5}$. Its targets
come from a CTC forced alignment of each utterance under the same chunk mask. Each token becomes visible to the decoder at
the end of the chunk that contains its last aligned frame plus $k$ frames, with $k$ drawn uniformly from
$\{1, 2, 3\}$. The decoder emits the wait token at the end of every chunk except the last. Batches mix single
utterances (0.425), utterances with full cross-attention (0.075), long sessions (0.30) and sessions restarted at random
points (0.20). A KL term with weight 0.25 pulls the streaming decoder towards the same decoder with full
cross-attention. The first stage used 215 GPU-hours on A100 GPUs.

\section{Inference server}
We export both models to ONNX and serve them with an open-source server built on ONNX Runtime \cite{onnxruntime}, which
runs on CPU or CUDA. For recorded audio, the server cuts each file at pauses into windows of at most 30\,s and collects
windows from concurrent requests into micro-batches, and one encoder session serves all three decoding setups. For live
audio, each WebSocket session caches the stream's encoder state, runs the encoder per chunk, and returns partial
transcripts followed by a final one.

A request can also supply a list of phrases, such as names of people and organizations, to favor during decoding. The
server compiles the list into an Aho-Corasick automaton \cite{aho1975efficient} over UTF-8 bytes, which needs no word
boundaries for Thai or Chinese. A hypothesis that follows a phrase gains a bonus for every matched character, withdrawn
if the match fails before the phrase is complete. CTC applies the bonus in a prefix beam search, and the offline decoder
adds it to its top-$k$ candidates at each step. The streaming decoder does not support word boosting yet.

\section{Experiments}
\subsection{Setup}
We evaluate on public test sets only. For Thai we use the six groups of a public Thai ASR benchmark:
FLEURS \cite{conneau2022fleurs} (read speech), an Isan dialect set \cite{isan}, TV talk shows, GigaSpeech~2
\cite{yang2024gigaspeech2} (online video), and the LOTUSDIS \cite{lotusdis} meeting corpus recorded with close and far
microphones. The benchmark summarizes the groups with a weighted CER, with weights 0.15, 0.15, 0.175, 0.175, 0.25 and
0.10 in that order. For English we use seven sets of the Open ASR Leaderboard \cite{srivastav2025openasr},
and for Chinese the AISHELL-1 \cite{bu2017aishell} test set, the WenetSpeech \cite{zhang2022wenetspeech} test\_net and
test\_meeting sets, and FLEURS.

We report CER for Thai and Chinese and word error rate (WER) for English, after the
Whisper English normalizer, traditional-to-simplified conversion with number normalization for Chinese, and the
benchmark's Thai normalizer. All Thai systems, including 13 commercial services queried in October and November 2026, ran through the same pipeline, as did Chinda ASR Pro v2 and Qwen-Audio 3.1 ASR Flash on English and Chinese. The offline Qwen-Audio 3.1 ASR Flash service returns an error instead of text for some clips, mostly short ones (4.8\% of LOTUSDIS clips and at most 1.7\% of any other set), and we score them as empty output. Other baselines are from one published source per table. Our training
data contains the train splits of FLEURS, the Isan set, GigaSpeech~2, LOTUSDIS, AISHELL-1, WenetSpeech, AMI, GigaSpeech,
SPGISpeech and VoxPopuli, so results on these test sets are in-domain. The training data of commercial services are not disclosed.

\subsection{Offline results}
Table~\ref{tab:thai} compares Thai results. \emph{Performance} (beam search) reaches 13.48\% weighted CER, 0.48 points behind
MAI-Transcribe-2 (13.00\%) and ahead of the other eight offline commercial services. \emph{Flash} (greedy CTC) reaches 15.23\%, lower than
all seven open offline models, of which Pathumma FastConformer is the best at 16.61\%. \emph{Performance} improves on \emph{Normal} (greedy decoding)
in every group, from 13.79\% to 13.48\% overall, a relative reduction of 2.2\%. \emph{Performance} has the lowest CER of all systems on the Isan set (4.79\% against 7.97\% for Pathumma FastConformer). On TV talk shows only Paxa STT lite is lower (4.57\% against 4.67\%). The Isan result benefits from the Isan train split. Meeting speech recorded 3 to 10\,m from the speaker
remains the main source of error, at 43.14\% CER for our best setup.

\begin{table}[t]
\centering
\footnotesize
\renewcommand{\arraystretch}{0.95}%
\fit{\begin{tabular}{llrrrrrrrr}
\toprule
& System & Params & Index & FLEURS & Isan & TV & GigaSpeech~2 & Close & Far \\
\midrule
API & MAI-Transcribe-2 & n/r & \textbf{13.00} & 3.94 & 13.12 & 5.24 & \textbf{8.60} & \textbf{16.22} & \textbf{39.69} \\
API & Paxa STT lite (preview) & n/r & 13.50 & 4.12 & 13.52 & \textbf{4.57} & 8.67 & 16.97 & 42.97 \\
API & Chinda ASR Pro v2 & n/r & 15.34 & 6.61 & 12.22 & 7.19 & 8.96 & 18.22 & 51.38 \\
API & Loma Sub & n/r & 17.42 & 5.54 & 10.12 & 5.03 & 9.75 & 23.22 & 66.84 \\
API & ElevenLabs Scribe v2 & n/r & 17.50 & 5.37 & 10.85 & 4.70 & 9.93 & 23.22 & 66.97 \\
API & Soniox v6 & n/r & 17.95 & 5.26 & 9.93 & 5.77 & 10.85 & 25.15 & 64.71 \\
API & Qwen-Audio 3.1 ASR Flash & n/r & 19.53 & \textbf{3.46} & 17.98 & 7.62 & 9.93 & 26.11 & 67.12 \\
API & Google Chirp 3 & n/r & 19.99 & 12.20 & 11.08 & 5.53 & 9.95 & 24.68 & 76.21 \\
API & Gemini 3.5 Flash-Lite & n/r & 22.68 & 5.32 & 19.80 & 8.27 & 14.24 & 30.37 & 73.80 \\
\midrule
Open & Pathumma FastConformer & 114M & 16.61 & 6.13 & 7.97 & 7.73 & 13.04 & 21.97 & 53.63 \\
Open & Typhoon Qwen-ASR 1.7B & 1.7B & 19.32 & 6.32 & 20.48 & 5.99 & 11.41 & 24.30 & 61.85 \\
Open & Typhoon Whisper turbo & 0.81B & 20.40 & 7.90 & 20.23 & 6.38 & 12.12 & 25.19 & 66.50 \\
Open & Lomasub-open & 166M & 21.14 & 8.66 & 18.74 & 5.50 & 11.62 & 25.72 & 76.03 \\
Open & Typhoon Whisper large-v3 & 1.55B & 21.40 & 7.43 & 12.21 & 5.72 & 12.08 & 27.56 & 84.50 \\
Open & Qwen3-ASR 1.7B & 1.7B & 22.86 & 6.55 & 27.44 & 10.07 & 11.96 & 26.33 & 73.20 \\
Open & Thonburian Whisper large-v3 & 1.55B & 25.29 & 7.05 & 27.49 & 12.32 & 16.30 & 29.53 & 77.22 \\
\midrule
Ours & \emph{Flash} (CTC) & 337M & 15.23 & 7.36 & 5.69 & 6.21 & 10.83 & 21.74 & 48.49 \\
Ours & \emph{Normal} & 337M & 13.79 & 6.66 & 4.84 & 4.77 & 9.64 & 20.52 & 44.16 \\
Ours & \emph{Performance} & 337M & 13.48 & 6.62 & \textbf{4.79} & 4.67 & 9.51 & 19.88 & 43.14 \\
\bottomrule
\end{tabular}}
\caption{Thai test sets, CER (\%). All systems ran in one pipeline. Index is the six-group weighted CER. n/r: not reported.}
\label{tab:thai}
\end{table}

On the archived Open ASR Leaderboard (Table~\ref{tab:en}), \emph{Normal} reaches an average WER of 7.53\%, lower than Whisper large-v3 (7.95\%) with less than a quarter of its parameters, and
behind Parakeet (6.44\%), Canary (6.81\%) and the commercial Qwen-Audio 3.1 ASR Flash (5.65\%). It does well on conversational and meeting speech, with 10.24\% on AMI
against 15.95\% for Whisper large-v3. It is weaker on read speech, with 3.42\% on LibriSpeech test-clean and 7.11\% on
test-other against 1.69\% and 3.19\% for Parakeet. \emph{Flash} is 1.83 points worse than \emph{Normal} on average
(9.36\% against 7.53\%), with the largest gaps on Earnings-22 (14.96\% against 12.02\%) and SPGISpeech (4.97\% against
2.44\%).

\begin{table}[t]
\centering
\footnotesize
\fit{\begin{tabular}{lrrrrrrrrr}
\toprule
System & Params & Avg & LS clean & LS other & AMI & Earnings-22 & GigaSpeech & SPGI & VoxPopuli \\
\midrule
Parakeet TDT 0.6B v2 \cite{parakeet2025tdtv2} & 0.6B & 6.44 & 1.69 & 3.19 & 11.16 & 11.15 & 9.74 & 2.17 & 5.95 \\
Canary 1B Flash \cite{zelasko2025canaryflash} & 1.0B & 6.81 & 1.48 & 2.87 & 13.11 & 12.77 & 9.85 & 1.95 & 5.63 \\
Whisper large-v3 \cite{radford2023whisper} & 1.55B & 7.95 & 2.01 & 3.91 & 15.95 & 11.29 & 10.02 & 2.94 & 9.54 \\
Whisper large-v3 turbo \cite{radford2023whisper} & 0.81B & 8.44 & 2.10 & 4.24 & 16.13 & 11.63 & 10.14 & 2.97 & 11.87 \\
Chinda ASR Pro v2 (API) & n/r & 8.03 & 1.78 & 3.42 & 19.38 & 11.21 & 9.94 & 2.98 & 7.52 \\
Qwen-Audio 3.1 ASR Flash (API) & n/r & 5.65 & 1.14 & 2.25 & 9.11 & 10.40 & 8.36 & 1.69 & 6.63 \\
\midrule
Ours, \emph{Flash} (CTC) & 0.34B & 9.36 & 4.70 & 9.04 & 11.15 & 14.96 & 11.40 & 4.97 & 9.27 \\
Ours, \emph{Normal} & 0.34B & 7.53 & 3.42 & 7.11 & 10.24 & 12.02 & 10.44 & 2.44 & 7.06 \\
Ours, \emph{Performance} & 0.34B & 7.47 & 3.34 & 6.87 & 10.21 & 12.07 & 10.45 & 2.42 & 6.95 \\
\bottomrule
\end{tabular}}
\caption{English test sets, WER (\%). Baselines are from the archived Open ASR Leaderboard \cite{srivastav2025openasr}
(December 2025), except Chinda ASR Pro v2 and Qwen-Audio 3.1 ASR Flash, which we ran in our pipeline. Avg is the mean over the seven sets shown, LS is LibriSpeech and SPGI is SPGISpeech. TED-LIUM is excluded because its references are no longer
distributed. n/r: not reported.}
\label{tab:en}
\end{table}

On Chinese (Table~\ref{tab:zh}), \emph{Normal} reaches 4.90\% CER on AISHELL-1 and 11.33\% on WenetSpeech
test\_net, behind Qwen-Audio 3.1 ASR Flash (1.34\% and 4.05\%), Paraformer \cite{gao2022paraformer} (1.95\% and 6.74\%) and SenseVoice \cite{funaudiollm2024}
(2.96\% and 7.84\%). On WenetSpeech test\_meeting it reaches 7.10\% (6.83\% with beam search), comparable to Paraformer and SenseVoice but behind Qwen-Audio 3.1 ASR Flash (2.59\%). Chinese made up 16\% of the
training mix, the smallest share of the three languages.

\begin{table}[b]
\centering
\begin{minipage}[t]{0.46\linewidth}\centering
\small
\fit{\begin{tabular}{lrrrrr}
\toprule
System & Params & AS-1 & WS net & WS mtg & FLEURS \\
\midrule
Paraformer-zh \cite{gao2022paraformer} & 0.22B & 1.95 & 6.74 & 6.97 & \\
SenseVoice-S \cite{funaudiollm2024} & 0.23B & 2.96 & 7.84 & 7.44 & \\
Whisper large-v3 \cite{radford2023whisper} & 1.55B & 5.14 & 10.48 & 18.87 & \\
Chinda ASR Pro v2 (API) & n/r & 5.08 & 8.63 & 14.17 & 8.45 \\
Qwen-Audio 3.1 ASR Flash (API) & n/r & 1.34 & 4.05 & 2.59 & 4.80 \\
\midrule
Ours, \emph{Flash} (CTC) & 0.34B & 4.91 & 11.39 & 7.48 & 10.64 \\
Ours, \emph{Normal} & 0.34B & 4.90 & 11.33 & 7.10 & 10.31 \\
Ours, \emph{Performance} & 0.34B & 4.62 & 10.61 & 6.83 & 10.01 \\
\bottomrule
\end{tabular}}
\caption{Chinese test sets, CER (\%). AS-1 is AISHELL-1 and WS is WenetSpeech. Published baselines are from the SenseVoice report
\cite{funaudiollm2024}, which does not report FLEURS. Chinda ASR Pro v2 and Qwen-Audio 3.1 ASR Flash were run in our pipeline. n/r: not reported.}
\label{tab:zh}
\end{minipage}\hfill
\begin{minipage}[t]{0.5\linewidth}\centering
\small
\fit{\begin{tabular}{llrrrr}
\toprule
Model & Decoding & LA (ms) & th & en & zh \\
\midrule
Offline & \emph{Flash} (CTC) & full & 15.23 & 9.36 & 8.61 \\
Offline & \emph{Normal} (decoder) & full & 13.79 & 7.53 & 8.41 \\
Streaming & \emph{Flash} (CTC) & 1,040 & 15.21 & 10.14 & 9.64 \\
Streaming & \emph{Flash} (CTC) & 320 & 16.26 & 10.96 & 10.38 \\
Streaming & \emph{Flash} (CTC) & 0 & 20.50 & 13.22 & 13.44 \\
Streaming & \emph{Normal} (streaming decoder) & 320 & 14.74 & 9.12 & 9.90 \\
\bottomrule
\end{tabular}}
\caption{Offline and streaming error rates with greedy decoding. LA: look-ahead. Thai is the weighted CER (\%), English
the average WER (\%) over the seven sets of Table~\ref{tab:en}, and Chinese the average CER (\%) over the four sets of
Table~\ref{tab:zh}. The streaming decoder runs as in a live session.}
\label{tab:stream}
\end{minipage}
\end{table}

\subsection{Streaming results}
\begin{figure}[t]
\centering
\includegraphics[width=\linewidth]{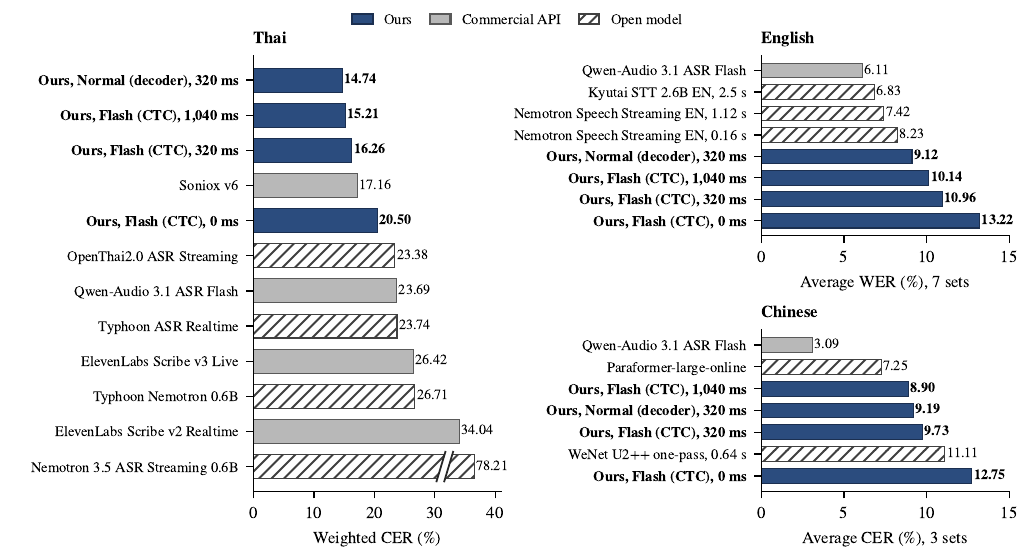}
\caption{Streaming results. Thai weighted CER, English average WER and Chinese average CER (\%) of our streaming setups
and other streaming systems (lower is better). All Thai systems ran in one pipeline, the commercial services as live
streams and the other open models on whole clips. ElevenLabs Scribe was asked to finalize once per clip. Nemotron 3.5 ASR Streaming
0.6B \cite{nvidia2026nemotron35} is not tuned for Thai, and its bar is cut. English and Chinese baselines are published
results, except Qwen-Audio 3.1 ASR Flash, which we ran as live streams. Averages cover the sets of
Figure~\ref{fig:thai}.}
\label{fig:stream}
\end{figure}

Table~\ref{tab:stream} compares the streaming model with the offline model, and Figure~\ref{fig:stream} compares it
with other streaming systems. With greedy CTC, the streaming model at 1,040\,ms look-ahead matches the offline model on
Thai (15.21\% against 15.23\% weighted CER), and shorter look-ahead raises the CER. The streaming decoder reaches 14.74\% at 320\,ms in a live session, 0.95 points above offline greedy decoding. Our 320 and 1,040\,ms setups are lower than every other streaming
system on Thai, including Soniox v6 (17.16\%) and OpenThai2.0 ASR Streaming \cite{iapp2026openthai} (23.38\%), also
trained on LOTUSDIS.

Streaming hurts English and Chinese more than Thai. At 1,040\,ms, greedy CTC is 0.78 WER points behind
offline CTC on English (10.14\% against 9.36\%) and 1.03 CER points behind on Chinese (9.64\% against 8.61\%). The
streaming decoder is 1.59 and 1.49 points behind offline greedy decoding (9.12\% and 9.90\%). Qwen-Audio 3.1 ASR Flash is more accurate on both languages, as are
Kyutai STT \cite{zeghidour2025dsm} and Nemotron Speech Streaming \cite{nvidia2026nemotronstream} on English and
Paraformer-large-online \cite{gao2022paraformer} on Chinese. WeNet U2++ \cite{zhang2022wenet2} trails our 320 and
1,040\,ms setups. Published baselines use their own text normalization.

\subsection{Inference cost}
\label{sec:cost}
Table~\ref{tab:cost} gives the accuracy and cost of each decoding setup. All run the full encoder and differ only in
decoding. We load-tested the released fp32 server with all setups loaded, following the procedure published with
the server, on one NVIDIA V100 32\,GB GPU and on 8 cores of an Intel Xeon Gold 5318Y. Requests were timed end to end, and
error rates at the highest load stayed within 0.08 points of one client.

\begin{table}[b]
\centering
\footnotesize
\fit{\begin{tabular}{lrrrrrrr}
\toprule
& Thai & \multicolumn{4}{c}{One V100 GPU} & \multicolumn{2}{c}{8 CPU cores} \\
\cmidrule(lr){3-6}\cmidrule(lr){7-8}
Setup & CER & xRT, 1 client & p50 (s) & xRT, saturated & p95 (s) & xRT, 1 client & xRT, saturated \\
\midrule
\emph{Flash} (CTC) & 15.23 & 212 & 0.04 & 637 & 1.96 & 61 & 64 \\
\emph{Normal} (greedy decoding) & 13.79 & 79 & 0.12 & $\ge$358 & 3.61 & 28 & 35 \\
\emph{Performance} (beam search) & 13.48 & 73 & 0.13 & 241 & 5.16 & 23 & 27 \\
\bottomrule
\end{tabular}}
\caption{Accuracy and cost of the decoding setups. Thai CER (\%) is from Table~\ref{tab:thai}, and speed is measured on
200 development clips (mean 9.45\,s). xRT is seconds of audio per second of wall time, with one client and at saturation
(128 clients on the V100, and 2, 16 and 8 on CPU for the three setups). The V100 p50 is the median latency with one client
and p95 the 95th-percentile latency with 128. \emph{Normal} did not saturate on the V100, so 358 is a lower bound.}
\label{tab:cost}
\end{table}

On the V100, \emph{Flash} reaches 2.6 times the saturated throughput of \emph{Performance} (637 against 241 times
real time) at a Thai CER 1.75 points higher. \emph{Normal} recovers 1.44 of those points at 358 times real time or more.
Beam search lowers the CER by a further 0.31 points at the cost of at least 32\% of that throughput. On CPU, one client
keeps most of the 8 cores busy.

The cost of live audio depends on the encoder look-ahead. A word appears only after its chunk (80, 400 or
1,120\,ms) has been processed, and the streaming decoder can wait longer. We streamed 2-minute Thai sessions in real
time, doubling the number of streams until the 95th-percentile lag exceeded the chunk length or the median lag grew over
the session. \emph{Flash} served 32 streams at 1,040\,ms and 16 at 320\,ms on both the V100 and
CPU. The streaming decoder (320\,ms) served 16 on the V100 and 8 on CPU, with the second CPU socket busy.
The true capacity lies between these counts and twice them. At these counts, the median per-stream 95th-percentile lag
was at most 0.040\,s on the V100 and 0.134\,s on CPU, and the median final transcript arrived within 0.08\,s (V100) and 0.17\,s (CPU) of the end of
the audio.

\subsection{Repetition before and after FTPO}
\label{sec:ftpo}
Table~\ref{tab:ftpo} compares greedy decoding of the pre-trained model with the released model, which adds the FTPO
fine-tune of Section~\ref{sec:training}. We count three kinds of failure over the 38,493 Thai test utterances.
\begin{itemize}
\item \textbf{Capped.} The output reaches the decoder's token limit, which we detect as 440 or more tokens when the
output is re-tokenized.
\item \textbf{Runaway.} The output is longer than twice its reference plus 10 characters.
\item \textbf{Loop.} The output is capped, or it repeats a unit of 4 to 60 characters three or more times in a row where
no reference does. Every capped output is therefore also a loop.
\end{itemize}

FTPO lowers the Thai weighted CER from 15.66\% to 13.79\%. Almost all of the gain is on meeting speech, where loops
were most frequent. Far-field CER falls from 55.07\% to 44.16\% and close-mic CER from 23.45\% to 20.52\%. Before the
fine-tune, the two meeting groups held 37 of the 40 capped outputs, 41 of the 50 runaways and 118 of the 134 loops.
After it, they hold 2, 4 and 47.

On the other test sets the counts barely move (3 to 1 capped, 9 to 8 runaways and
16 to 13 loops), and their weighted CER changes from 6.60\% to 6.53\%. Within them, Isan rises by 0.26 points. A
control fine-tune with the same steps and data but without the FTPO loss shows the same rise on held-out Isan data, so
we attribute it to the extra training rather than to FTPO. Greedy CTC is unchanged (15.23\%) because the encoder and
the CTC head are frozen. On far-field speech, greedy decoding goes from worse than greedy CTC (48.49\%) to better.

\begin{table}[t]
\centering
\footnotesize
\begin{tabular}{lrrrrrrrrr}
\toprule
& & \multicolumn{2}{c}{CER (\%)} & \multicolumn{2}{c}{Capped} & \multicolumn{2}{c}{Runaway} & \multicolumn{2}{c}{Loop} \\
\cmidrule(lr){3-4}\cmidrule(lr){5-6}\cmidrule(lr){7-8}\cmidrule(lr){9-10}
Group & Utterances & Pre & Post & Pre & Post & Pre & Post & Pre & Post \\
\midrule
Non-meeting & 10,913 & 6.60 & 6.53 & 3 & 1 & 9 & 8 & 16 & 13 \\
Near-field mic & 16,548 & 23.45 & 20.52 & 12 & 2 & 13 & 2 & 44 & 28 \\
Far-field mic & 11,032 & 55.07 & 44.16 & 25 & 0 & 28 & 2 & 74 & 19 \\
\midrule
All & 38,493 & 15.66 & 13.79 & 40 & 3 & 50 & 12 & 134 & 60 \\
\bottomrule
\end{tabular}
\caption{Thai test results of greedy decoding before (Pre) and after (Post) the FTPO fine-tune, per group. Non-meeting
covers FLEURS, the Isan set, TV talk shows and GigaSpeech~2, and its CER is weighted as in the index. Near-field and far-field
mic are LOTUSDIS meetings on close microphones and on microphones 3 to 10\,m away. The All row gives the weighted CER and the total counts. Capped,
Runaway and Loop count the failures defined in Section~\ref{sec:ftpo}. The pre-trained model was decoded in fp16
and the released model in fp32. On development data, the two precisions agree within 0.05 CER.}
\label{tab:ftpo}
\end{table}

\section{Related work}
Large multilingual recognizers trained on weakly or pseudo-labeled data, such as Whisper \cite{radford2023whisper},
Canary and Parakeet \cite{sekoyan2025canary} and Qwen3-ASR \cite{qwen3asr}, set the accuracy standard for English and Chinese. FireRedASR \cite{xu2025fireredasr} is a large recognizer for Chinese trained mostly on human-transcribed Mandarin. Most use an attention or language-model decoder, with flagship
checkpoints from 0.6B to over 8B parameters, while SenseVoice \cite{funaudiollm2024} and Paraformer
\cite{gao2022paraformer} reach high Chinese accuracy non-autoregressively. Our model is smaller than most and keeps a
CTC head next to the decoder, so the cheapest setup needs no autoregression.

Open Thai recognizers are mostly fine-tuned versions of Whisper, such as Thonburian Whisper \cite{thonburian}, the
Typhoon Whisper models \cite{typhoonwhisper} and the distilled Lomasub-open \cite{lomasub2026open}, together with FastConformer models from Pathumma \cite{pathumma} and Typhoon
\cite{typhoon2026rt,typhoon2026stream}. Our model covers all three languages in one checkpoint, with offline and streaming variants.

Cache-aware streaming FastConformer \cite{noroozi2024stateful} restricts attention to chunks with cached left context and
trains one model for several look-aheads. Our streaming encoder follows this design. Contextual biasing by shallow fusion
with a prefix tree or automaton is common for CTC and transducer models \cite{biasctc,biasrnnt}, and our server
applies one byte-level automaton to CTC and to the offline decoder.

\section{Conclusion}
We presented VISAI OpenSpeech, a 337M-parameter FastConformer model with a CTC head and an attention decoder for Thai,
English and Chinese, and its streaming variant. With beam search it reaches 13.48\% weighted CER on a six-group Thai benchmark, behind
one of the commercial services and ahead of every open model we evaluated. From the same checkpoint on one V100 GPU under load, greedy CTC
runs at 637 times real time, while beam search runs at 241 times and lowers Thai CER by 1.75 points. The streaming variant matches offline CTC at
1,040\,ms look-ahead.

\clearpage
\section*{Limitations}
The current models and evaluation have the following limitations.
\begin{itemize}
\item \textbf{Repetition.} The attention decoder can still repeat itself on difficult audio. After FTPO, 60 of the
38,493 Thai test outputs under greedy decoding contain a loop, including 3 that reach the token limit
(Table~\ref{tab:ftpo}). We have not measured loops for beam search, the streaming decoder, English or Chinese. Greedy
CTC cannot loop, so \emph{Flash} is the safer setup for noisy or long recordings. On development data, a second round of FTPO on 577 newly mined pairs lowers loops on 5,236 far-field utterances from 9 to 6 after 1,000 steps but raises the weighted Thai CER from 16.97\% to 17.35\%. A control fine-tune without FTPO reaches 18.07\% with 38 far-field loops, so extra training hurts CER and FTPO only partly offsets the rise.
\item \textbf{Far-field speech.} Thai meeting speech recorded 3 to 10\,m from the speaker remains the hardest
condition, with a CER between 43.14\% and 48.49\% across the three setups (Table~\ref{tab:thai}).
\item \textbf{Word boosting.} We have not measured how much word boosting improves the recognition of names and rare
phrases, how often it inserts a boosted phrase that was not spoken, or its speed overhead. The streaming decoder does
not support it yet.
\item \textbf{Inference cost.} We measured cost only for our fp32 server on one V100 GPU and 8 CPU cores, on
development clips rather than the test sets, one decoding setup at a time and with the language given. We have not measured fp16
engines, streaming at 0\,ms look-ahead, or the speed of other models and services.
\item \textbf{Prompt options.} Our test scores are computed on normalized text, so they do not evaluate written or spoken
forms of numbers, dates and units (ITN) or punctuation and case (PNC). We have not measured ITN, PNC or segment
timestamp accuracy, nor decoder test accuracy when the language tag is left out.
\item \textbf{Short clips and numbers.} On clips shorter than about 2\,s, greedy CTC can return empty text or
characters of another language. Thai numbers read one digit at a time, such as a year read as ``five four'', can come
out as a spoken cardinal (``fifty-four''), because the training transcripts spell out every Thai integer as a cardinal. As a
result, the ITN flag does not always behave as expected for numbers. We intend to fix both in the next release.
\end{itemize}

\section*{Acknowledgements}
We thank the Vidyasirimedhi Institute of Science and Technology (VISTEC) for sponsoring the compute used in this
work.

\bibliography{references}
\fcend

\end{document}